\documentclass[a4, conference]{IEEEtran}
 \IEEEoverridecommandlockouts
 \usepackage{cite}
 \usepackage{amsmath,amssymb,amsfonts}
 \usepackage{graphicx}
 \usepackage{textcomp}
 \usepackage{xcolor}
 \usepackage{dblfloatfix}    
 \usepackage{eso-pic} 
 \AddToShipoutPicture*{\small\sffamily\raisebox{1.6 cm}{\hspace{2 cm}*For citation please use the latest/final version of this paper: https://ieeexplore.ieee.org/document/9221525}}

\def\BibTeX{{\rm B\kern-.05em{\sc i\kern-.025em b}\kern-.08em
  T\kern-.1667em\lower.7ex\hbox{E}\kern-.125emX}}

 \usepackage{multirow}
 \usepackage{eso-pic} 
 \usepackage{multirow}

 \usepackage{booktabs}
 \usepackage{footnote}
 \usepackage{etoolbox}
 \usepackage{amssymb}

\usepackage{paralist}

\usepackage[T1]{fontenc}
\usepackage{algorithm,algpseudocode,amsmath}
\usepackage{slashbox}
\usepackage{mathtools}
\usepackage{pgfplots}
\usepackage{pgfplotstable}
\usepackage{xcolor}

\usepackage{soul}
\usepackage{tikz}
\usetikzlibrary{calc}
\usepackage{relsize}
\usepackage{graphicx, caption, subcaption}
\usepackage{cleveref}
\usepackage[moderate,tracking=normal, title=normal]{savetrees}
\usepackage{float}
\usepackage{tkz-euclide}
\usepgfplotslibrary{fillbetween}
\usepackage{amsmath}

\usetikzlibrary{arrows, arrows.meta, patterns}
\usetikzlibrary{decorations.pathreplacing,calc}
\pgfplotsset{compat=1.10}
\usepackage{soul}
\usepackage{xcolor}
\usepackage{listings}

  \definecolor{mGreen}{rgb}{0,0.6,0}
  \definecolor{mGray}{rgb}{0.5,0.5,0.5}
  \definecolor{mPurple}{rgb}{0.58,0,0.82}
  \definecolor{backgroundColour}{rgb}{0.95,0.95,0.92}
  \usepackage{listings}
  \definecolor{mygray}{gray}{0.6}

  \newcounter{phase}[algorithm]
  \newlength{\phaserulewidth}
  \newcommand{\setphaserulewidth}{\setlength{\phaserulewidth}}
 
 \makeatother

\setphaserulewidth{.7pt}

\makeatletter                   
\def\footnoterule{\relax%
  \kern-5pt
  \hbox to \columnwidth{\hfill\vrule width 1\columnwidth height 0.4pt\hfill}
  \kern4.6pt}

\makeatother
\usepackage{color,colortbl}
\definecolor{MyGray}{gray}{0.8}
\usepackage{float}        
\usepackage{multirow}
\usepackage[T1]{fontenc}       
\newcommand{\ditto}[1][.4pt]{\textquotedbl}

\newcommand{\rsec}[1]{Section\,\ref{#1}}
\newcommand{\rfig}[1]{Fig.\,\ref{#1}}
\newcommand{\rtab}[1]{Tab.\,\ref{#1}}
\newcommand{\requ}[1]{(\ref{#1})}

\begin{document}
\title{HyperLogLog Sketch Acceleration on FPGA}
\author{
\IEEEauthorblockN{Amit Kulkarni, Monica Chiosa, Thomas B. Preu{\ss}er, Kaan Kara, David Sidler and Gustavo Alonso}
\IEEEauthorblockA{Systems Group, Department of Computer Science, ETH Zurich\\
Universit{\"a}tstrasse 6, 8092 Zurich\\
}}
\maketitle
\begin{abstract}
Data sketches are a set of widely used approximated data summarizing techniques. Their fundamental property is sub-linear memory complexity on the input cardinality, an important aspect when processing streams or data sets with a vast base domain (URLs, IP addresses, user IDs, etc.). Among the many data sketches available, HyperLogLog has become the reference for cardinality counting (how many distinct data items there are in a data set). Although it does not count every data item (to reduce memory consumption), it provides probabilistic guarantees on the result, and it is, thus, often used to analyze data streams. In this paper, we explore how to implement HyperLogLog on an FPGA to benefit from the parallelism available and the ability to process data streams coming from high-speed networks. Our multi-pipelined high-cardinality HyperLogLog implementation delivers $1.8\times$ higher throughput than an optimized HyperLogLog running on a dual-socket Intel Xeon E5-2630~v3 system with a total of 16 cores and 32 hyper-threads.
\end{abstract}
\section{Introduction}
Calculating basic statistics over large data collections is the first step in many data analytic procedures. Either as the direct result of user queries or as an initial step for other, more complex operations on the data, computing the cardinality, item frequency, distribution, heavy-hitters, or top-K elements are nowadays standard operations in both data streaming as well as over distributed data processing engines. Item frequency, for instance, is essential to identify from a data stream recording web accesses how often individual web pages are accessed. Frequent items~\cite{freq_item} are used to identify, e.g., the users that most frequently request a given service. Similarly, cardinality is used to, e.g., determine how many different users are utilizing a given service or how many distinct items are being bought from an e-shop given a list/stream of accesses or purchases. 

Common to all these operations is the problem of space complexity. If the domain from where the data set is derived is very large (IP addresses, URLs, user IDs, items available from a catalog), a naive approach to counting becomes linear on the cardinality of the data set, and that might involve potentially millions or even billions of entries to keep track of. As a result, existing systems resort instead to approximation through several techniques collectively referred to as \textit{sketch algorithms}~\cite{cormode}. These algorithms only provide estimates, rather than accurate counts, for a variety of statistics on the data, but they do so using a fixed amount of space. The algorithms provide well defined analytical bounds on the precision that will be reached as a function of the space used to perform the actual operation. These bounds involve relatively low error margins that are often acceptable when processing large data sets.

In this paper, we focus our attention on the cardinality problem, i.e., how to determine the number of distinct elements on a data collection. This function can be found in many data processing systems and network monitoring applications~\cite{heule}. For instance, in SQL, it is used to implement \textit{COUNT(DISTINCT ...)} that returns the number of distinct items in a column. Of the different ways to estimate the cardinality of a multiset, HyperLogLog (HLL) is nowadays the standard algorithm~\cite{flajolet}. Google, for instance, uses HLL in BigQuery, a cloud-based distributed data processing system, to estimate the cardinality of data sets with several billion distinct items with errors lower than 1\%~\cite{google_blog}. 

In modern cloud and data center environments, the widely adopted separation of compute and storage often means that an operation such as calculating the cardinality of a data set involves reading the data from storage and forwarding it to the computing nodes. It follows that having the ability to perform HLL over streams of data would be very beneficial, especially if it can be done close to the network without involving the CPU, and avoiding expensive copying of the data to memory. With this in mind, we explore the implementation of HLL on an FPGA to benefit from both the inherent parallelism as well as its architectural flexibility. In the paper, we describe the implementation of HLL on an FPGA, how to parallelize it using multiple concurrent pipelines, and how it can be embedded in an FPGA-based Network Interface Card (NIC) supporting TCP/IP to perform the cardinality estimation directly on the network. The main contributions of the paper include: 

\textbf{1)} A single-pipelined dataflow architecture implementing HLL on an FPGA with a performance improvement of $2\times$ over the throughput reachable by a single-threaded CPU implementation. 

\textbf{2)} A multi-pipelined parallel architecture of HLL with a performance improvement of $1.8\times$ compared to the 16-cores, 32-thread CPU implementation.

\textbf{3)} A design embedding the HLL implementation on a NIC so that it can process data streams arriving through a 100\,Gbit/s TCP/IP link. This design highlights the advantage of network-faced processing with FPGAs compared to doing the same operation on a CPU, where it becomes compute bound.

The rest of the paper is organized as follows: \rsec{sec:relatedwork} provides an overview of background and related work. A detailed explanation of the HyperLogLog algorithm is presented in \rsec{sec:hll}. We profiled HLL with different data sets to decide on the parameters for hardware implementation. Comprehensive details on the profiled results are described by \rsec{sec:profiling}. \rsec{sec:hll_df} presents the proposed hardware architectures of HLL. The experiments and the results are discussed in \rsec{sec:exp} followed by describing the network integration of HLL in \rsec{sec:tcp}. Finally, \rsec{sec:conc} concludes the paper.
\section{Background and Related Work}
\label{sec:relatedwork}
This work combines ideas from diverse fields of study: data sketch algorithms, specialized hardware solutions, and in-network data processing.

\subsection{Cardinality Estimation}
Data sketch algorithms play an essential role in big data. Their goal is to obtain approximate, yet accurate statistics about the data with sublinear time or space complexity, or both. The statistics gathered by sketches can be used in approximate query processing~\cite{chaudhuri2017approximate} and are fundamental to perform query optimization~\cite{youssefi1979query,zhang2009kernel,neumann2011characteristic} in database management systems (DBMS).
Cardinality estimation is a family of sketching techniques that is often used in these mentioned scenarios, besides also being utilized in other diverse application areas such as network security~\cite{estan2003bitmap}, network size estimation~\cite{varagnolo2013distributed}, and data mining~\cite{metwally2008go}. 
HyperLogLog~\cite{flajolet} is known to be one of the best algorithms, achieving high accuracy over all cardinality ranges and being trivially parallelizable, leading to efficient scale-out implementations~\cite{heule}.

\subsection{Specialized Hardware for Data Processing}

Due to stagnating single-core performance in CPUs and the recent slowdown in technology scaling (described as a slowdown in Moore's Law~\cite{eeckhout2017moore}), specializing hardware has become widespread to achieve high performance and efficiency in data processing systems. Prominent examples include Microsoft's and Amazon's deployment of FPGAs in their datacenters~\cite{putnam2014reconfigurable,amazonf1}, Google's development of a specialized processor called Tensor Processing Unit~\cite{jouppi2017datacenter}, and Intel embedding FPGAs next to Xeon CPUs in the same package with a coherent interconnect~\cite{oliver2011reconfigurable}.

Specialized hardware solutions to accelerate relational processing operators such as joins~\cite{halstead2013accelerating,kara2017fpga}, aggregation~\cite{mueller2009data}, and sorting~\cite{mashimo2017high} show increased performance and efficiency. István et al.~\cite{istvan2014histograms} show that FPGAs can be used to build histograms without affecting the throughput. 
Kara et al.~\cite{hash_accel} show that robust and expensive hash functions can be implemented on an FPGA without any reduction in the processing rate, as opposed to performing these hash functions on a CPU. Tong et al.~\cite{mc1,mc11} show Count-Min sketch acceleration and its application in high-speed networks.

Cardinality estimation algorithms can also be performed on the data path and most popular algorithms in this domain require robust hashing. 
Consequently, the FPGA-based implementation of HyperLogLog presented in this paper utilizes similar ideas but expands upon them by presenting an end-to-end cardinality estimation solution.

\subsection{In-network Data Processing}
As the datacenter network bandwidth keeps increasing, with 100\,Gbit/s recently becoming the norm, the burden on CPUs to process data as quickly escalates. Besides offloading network processing to network interface card (NIC), there are also ongoing efforts to offload parts of the application logic to NICs to reduce the data processing burden of CPUs. Standalone FPGA platforms can also be used as NICs, thanks to efforts in developing FPGA-based TCP/IP~\cite{sidler2015scalable} or RoCE~\cite{sidlerrepo} stacks. In an FPGA-based NIC, the remaining logic on the FPGA chip can be used to perform complex data processing tasks on the received or transmitted data.
For instance, Microsoft uses this setup in their datacenters to, e.g., accelerate the Bing search engine~\cite{putnam2014reconfigurable} or perform low-latency neural network inference~\cite{chung2018serving}.

Creating sketches as the data is received from the network can be useful for multiple reasons: (1) The FPGA is better suited to perform complex data processing tasks with high throughput. It can match the 100\,Gbit/s line rate when creating the sketch. (2) The FPGA creates the sketch as the data is received, it is practically ``for\,free''. This frees up the CPU to perform other tasks. We integrate our FPGA-based HLL implementation next to an FPGA-based TCP/IP stack to estimate the cardinality at line-rate. 
\vspace*{-2pt}\section{HyperLogLog}~\label{sec:hll}
The HyperLogLog (HLL) sketch is used to determine the cardinality of large multisets without the necessity of storing every data item. It uses hash-based data randomization to approximate this cardinality. The algorithm monitors the maximum leading zero counts of the encountered hash values. Hash values with more leading zeros are less likely. So, their observation indicates a higher cardinality. The hashing algorithm is assumed to produce uniformly distributed hash values.

\begin{table}
\centering
\caption{4-bit hash values.}
\label{tab:hash_values}
\begin{tabular}{cccccccc}\toprule
\underline{0000} & \underline{000}1 & \underline{00}10 & \underline{00}11 &
\underline{0}100 & \underline{0}101 & \underline{0}110 & \underline{0}111\\
1000 & 1001 & 1010 & 1011 & 1100 & 1101 & 1110 & 1111\\\bottomrule
\end{tabular}
\end{table}
To understand HLL, assume that the hash function randomly draws the binary representation of a number from the range $[0:15]$ as listed in \rtab{tab:hash_values}. If every hash value occurs with the same probability, we observe the following: the probability that the hash has at least one leading zero is 50\% (8/16); the probability of the hash having at least two leading zeros is 25\% (4/16); the probability of the hash containing at least three leading zeros is 12.5\% (2/16), and the probability of the hash containing four leading zeros is 6.25\% (1/16). Statistically speaking, around 8~elements hashing to different values are needed before encountering a hash starting with three leading zeros. Generalizing, we need to see around $2^k$ elements to observe a hash containing $k$ leading zeros. Conversely, if a maximum of $k$ leading zeros has been seen, one has probably processed $2^k$ different elements. This is the main intuition behind HLL. However, if based on a single measurement, this approach results in a large variance of the estimated cardinality. This is the case, e.g., when we observe a large number of leading zeros in a hash value very early.

The estimation variance, and hence the expected estimation error, is reduced by stochastic averaging~\cite{stochastic}. For this purpose, the hash value is divided into two parts: (a) a short bucket index\,$i$ and (b) a remaining hash\,$w$. The bucket index splits the stream into disjoint substreams. For each one of them, a designated counter $M[i]$ is maintained that tracks the maximum rank $\varrho(.)$ so far observed in the associated substream of hashes where the rank $\varrho(w)$ is the number of leading zeros in $w$ plus one. This approach of averaging observations across buckets reduces the error that random occurrences produce.

At any point in time, the cardinality can be estimated by taking the harmonic mean of the estimates implied by the individual bucket ranks. This estimate comes with a standard error. The HLL standard error is $\frac{1.04}{\sqrt{m}}$ with a space complexity of $O(\epsilon^{-2} \log\log n + \log n)$ in the data streaming ($\epsilon,\delta$)-model~\cite{indyk, Kane}, where $\epsilon$ is the confidence parameter, $\delta$ is the approximation parameter of the data stream, and $m$ is the number of buckets.

\begin{figure}
\resizebox{\linewidth}{!}{\includegraphics[trim=4.6cm 4.85cm 4.85cm 5cm,clip=true, width=0.8\linewidth]{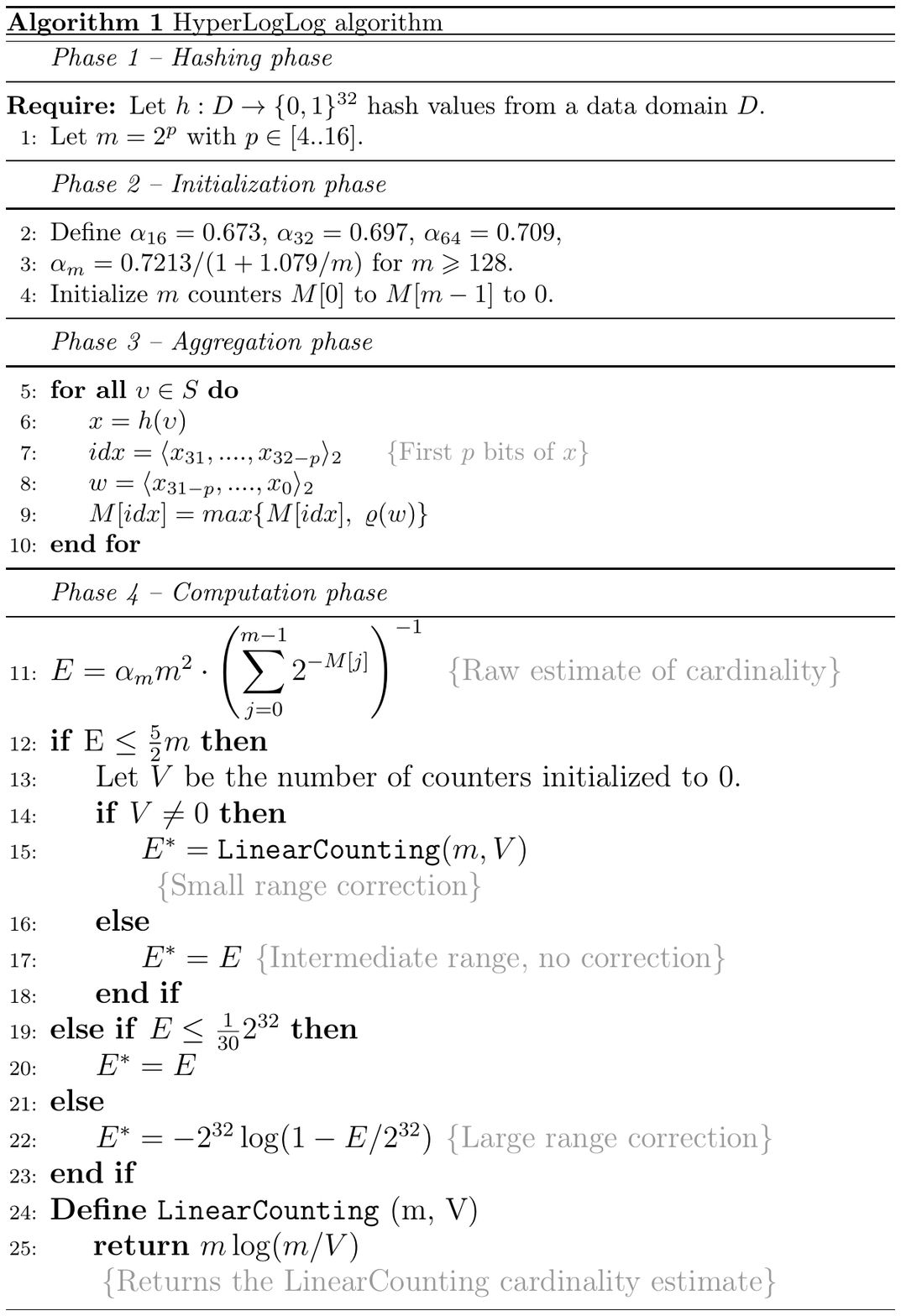}}
\end{figure}

A practical variant of the original HLL~\cite{flajolet} is shown as Algorithm~1. It has four phases:

\textbf{1) \textit{Hashing}}: Every data item from the data stream (multiset) is hashed using a hash function that produces a 32-bit hash value.
 
\textbf{2) \textit{Initialization}}: Depending on the chosen value of $p\in\left[4:16\right]$, a constant $\alpha_{m}$ with $m=2^p$ is calculated as listed in line~3 of Algorithm~1. This constant will be used for bias correction. The array $M\left[0:m-1\right]$ of bucket counters is initialized to all $0$.
 
 \textbf{3) \textit{Aggregation}}: The first $p$ bits of each hash value act as an index\,$i$ to divide the data stream\,$S$ into\,$m$ substreams\,$S_i$. They identify the associated bucket counter\,$M[i]$. The remaining bits\,$w$ of the hash value are subjected to the rank computation\,$\varrho(w)$. The associated bucket counter will be updated to the maximum of its current value and this newly determined rank. Ultimately, this yields:
\begin{align}
M[i] = \max_{w \in S_i }\varrho(w)
\label{eqn1}
\end{align}

\textbf{4) \textit{Computation}}: A raw cardinality estimate is obtained as the product of the harmonic mean of the individual substream cardinality estimates $2^{M[i]}$, the substream count $m$, and the bias correction $\alpha_m$. To compensate a systematic overestimation of small cardinalities, the algorithm reverts to \texttt{LinearCounting} in these cases as shown in line 15. Intermediate ranges of cardinalities require no correction (line 17). However, when the data set approaches large cardinalities on the order of $10^9$, the 32-bit hash function is increasingly unable to differentiate data items due to hash collisions. The correction of line~22 tries to mitigate this effect. 

A hash function producing $H$-bit hash values can distinguish at most $2^H$ data inputs. So, hash collisions are imminent if the cardinality of the data set approaches $2^H$. Fortunately, the memory footprint of HLL does not grow linearly with $H$. The memory requirements for HLL are determined by the number of buckets and the maximum rank $\varrho(.)$. For an $H$-bit hash function and a precision $p$, we obtain:
\begin{align}
\varrho(.) &\le H-p+1 && \mbox{\textcolor{mygray}{\small\{Maximum observable rank\}}}\\
B &= 2^p\cdot\log_2\left(H-p+1\right) && \mbox{\textcolor{mygray}{\small\{Memory footprint in bits\}}}\label{eqn3}
\end{align}
Switching from a 32-bit to a 64-bit hash function boosts the accuracy of estimating large cardinalities significantly. Typically, 64-bit hash functions are used to estimate multisets of cardinalities beyond 1 billion~\cite{heule}. This modification increases the necessary size of each counter by one bit only. Note that this choice renders the large range correction obsolete for all conceivable practical multiset cardinalities.

The accuracy benefits of choosing a 64-bit hash function are obtained at the cost of an approximately doubled compute effort as compared to a 32-bit hash function. In a compute-bounded setting, as on a CPU, this translates into a corresponding reduction of the processing rate. On the other hand, the HLL algorithm is structurally simple enough to quickly become I/O-bound on an FPGA platform. Hence, the accuracy gain can be easily realized by expanding in fabric space rather than by sacrificing throughput. In the next section, we profile the HLL algorithm from a statistical perspective, showing the necessity for a 64-bit hash when approaching high cardinalities.
\section{HyperLogLog Profiling}\label{sec:profiling}
\begin{figure}
  \begin{subfigure}[b]{.5\linewidth}
  \centering
    \includegraphics[trim=6.7cm 10.2cm 6.5cm 10.3cm,clip=true, width=\linewidth]{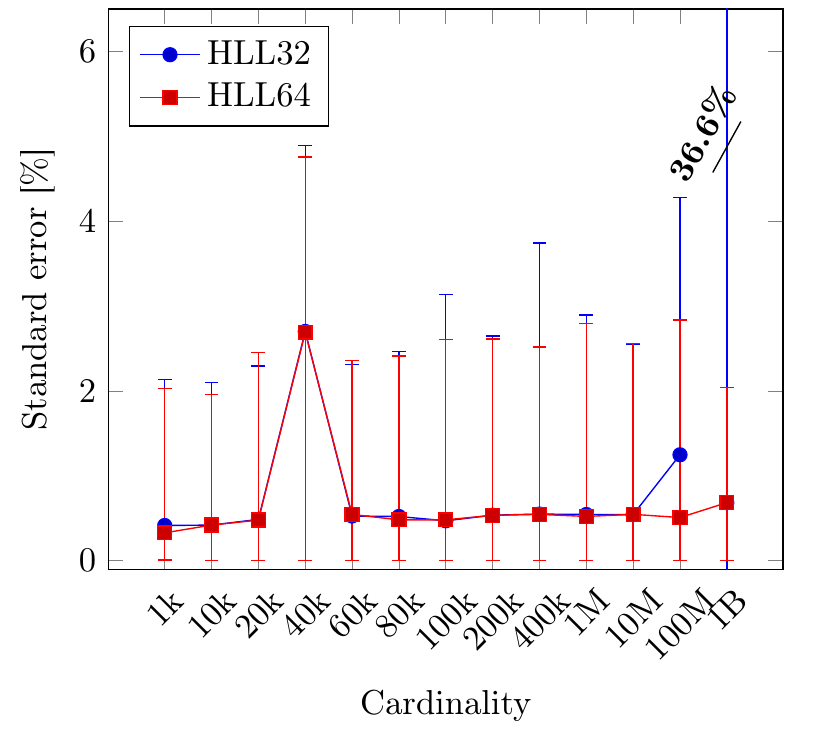}
    \vspace*{-6pt}
     \caption{\scriptsize p=14}
    \label{fig:p=14} 
  \end{subfigure}%
  \begin{subfigure}[b]{.5\linewidth}
  \centering
    \includegraphics[trim=6.7cm 10.2cm 6.5cm 10.3cm,clip=true, width=\linewidth]{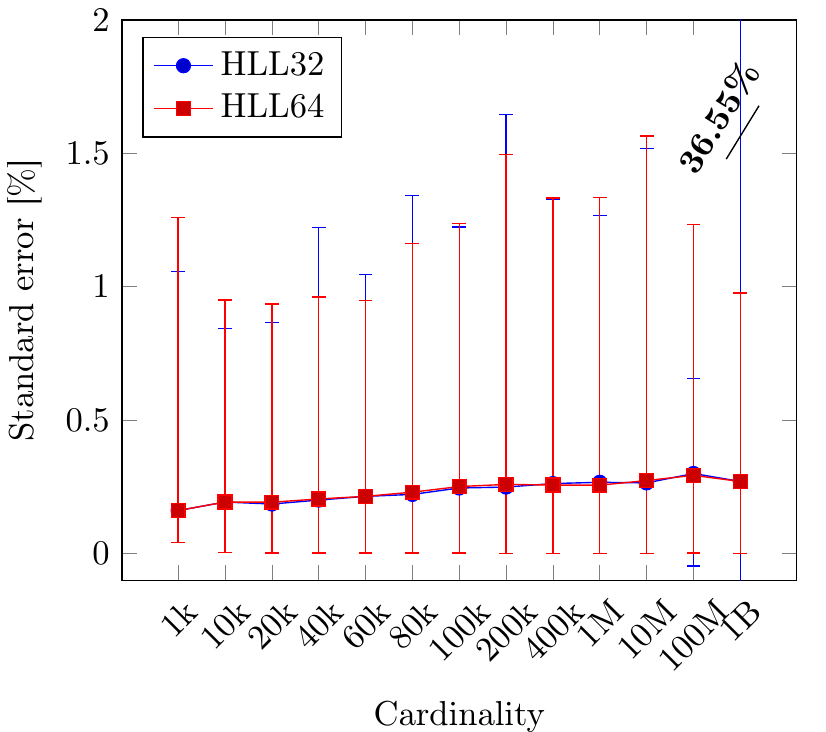}
    \vspace*{-6pt}
    \caption{\scriptsize p=16}
    \label{fig:p=16} 
  \end{subfigure}
 \caption{HyperLogLog standard error.}
\label{fig:hll_profile} 
\end{figure}

\begin{figure*}
\begin{center}
\includegraphics[trim=0.6cm 0.65cm 0.6cm 0.8cm,clip=true, width=0.7\linewidth]{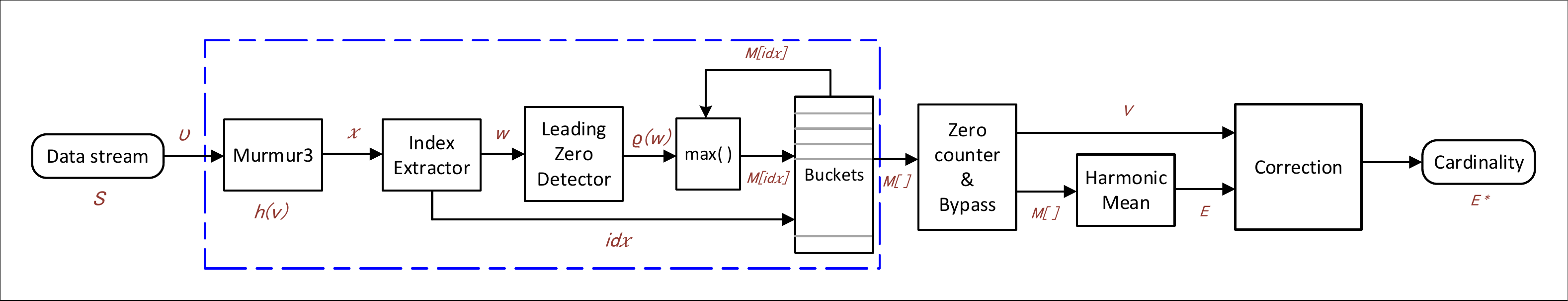}
\caption{A single-pipelined HyperLogLog dataflow engine.}
\label{fig:hll_single}
\end{center}
\end{figure*}

We profile the HLL algorithm to evaluate its statistical properties. This profiling is independent from any particular platform-specific implementation. We explore the parameter space $(p, H)\in\{14,16\}\times\{32,64\}$ using synthetic data sets. The data sets are generated by randomly sampling the range $[0:2^{32}-1]$.  The input sequence is hashed using the Murmur3 hash function \cite{mumur} of the respective bit width.

The standard error for HLL (of different hash sizes) with $p=14$ is depicted in \rfig{fig:p=14}. We consider the maximum, minimum, and median of the standard errors obtained from multiple data points. Clearly, the 32-bit hash HLL estimates the cardinality within reasonable error bounds for cardinalities up to $10^8$. For data sets with larger cardinalities, the standard error quickly grows beyond 30\%, which is often not acceptable anymore. HLL reverts to \texttt{LinearCouting} for cardinalities below a threshold of $\frac{5}{2}\cdot m$. The transition between the estimation schemes  occurs at about $40k$ for $p=14$. This location is identified by a local increase of the observed maximum estimation error of up to 5\%. These error curves are similar to the results reported by Heule et al.~\cite{heule}.

To sustain the cardinality estimation beyond $10^8$, we increased the hash size from 32 to 64~bits. This results in a significant reduction of the standard error. To reduce the standard error further, we increased the precision to $p=16$. \rfig{fig:p=16} shows the corresponding standard error variation. A 32-bit hash achieves a standard error less than 2\% for all data sets of a cardinality below $10^8$. However, the standard errors surge quickly above 35\% beyond this point. In the case of the 64-bit hash, the standard error remains close to 1\% for the whole cardinality range. It is to be noted that the theoretical average standard error of HLL is given by $\frac{1.04}{\sqrt{m}}$. With $p=16$, the expected standard error is $0.41\%$. The average standard error of our experiments shown in \rfig{fig:hll_profile} stays below this expected average.

\begin{table}
\caption{HyperLogLog memory footprint.}
\label{tab:mem}
\centerline{\begin{tabular}{lcccc}\toprule
$p$ [bits] & 
\multicolumn{2}{c}{14} &
\multicolumn{2}{c}{16} \\
$H$ [bits] & 32 & 64 & 32 & 64 \\\midrule
\begin{tabular}[c]{@{}c@{}}$\lceil \log_2 (H - p + 1) \rceil$ \\ register size [bits]\end{tabular}
  & 5 & 6 & 5 & 6 \\\midrule
Total memory {[}KiB{]} & 10 & 12 & 40 & 48  \\\bottomrule
\end{tabular}
}
\end{table}

\rtab{tab:mem} summarizes the memory requirements for the explored parameter settings as obtained from \requ{eqn3}. Observe that the transition from a 32-bit to 64-bit hash only implies a 20\% increase of the memory footprint. Increasing the precision $p$ from $14$ to $16$, on the other hand, quadruples the partitioning into buckets effecting a corresponding growth in counter memory.

Having attained the best accuracy results for $p=16$ and a $64$-bit Murmur3 hash function, we describe this configuration for our HLL implementation on an FPGA.

\section{HyperLogLog Hardware Architecture}\label{sec:hll_df}

\begin{figure*}[t]
\begin{center}
\includegraphics[trim=0.6cm 0.85cm 0.6cm 0.9cm,clip=true, width=0.7\linewidth]{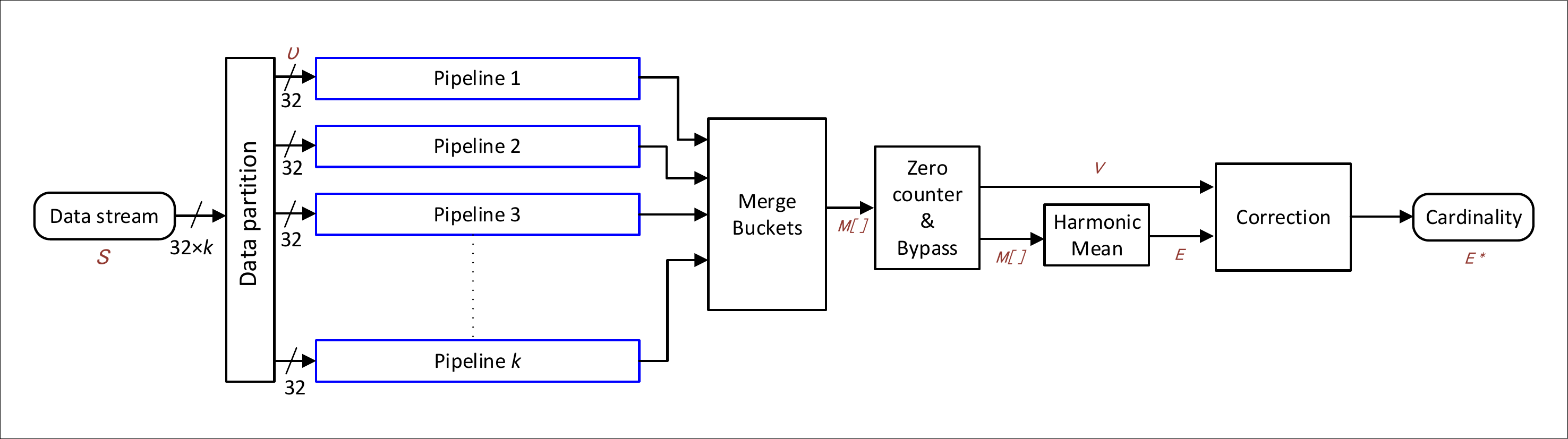}
\caption{A multi-pipelined parallel HyperLogLog dataflow architecture.}
\vspace{-15pt}
\label{fig:hll_parallel}
\end{center}
\end{figure*}

\subsection{Dataflow Architecture}
\rfig{fig:hll_single} depicts a single pipelined dataflow engine for the HLL sketch on the FPGA. The annotations in the figure identify the corresponding variables of Algorithm~1. 

The design is implemented in C++ using the high-level synthesis (HLS) using Vivado~HLS 2019.1 with a target frequency of 322\,MHz and an initiation interval of $II=1$. The initiation interval is defined as the number of clock cycles between successive data inputs to the pipeline~\cite{hls_book}.

The pipeline structure encompassed in blue dashes in \rfig{fig:hll_single} forms the aggregation phase of the HLL algorithm. It processes the incoming data, received from an AXI4 stream interface, tracking the maximum ranks of the substreams in on-chip Block RAM (BRAM). Once all data items have been processed, the buckets module starts forwarding the counter values. This marks the hand-over to the computation phase of the HLL algorithm.

\subsubsection{\textbf{Hash Function}}
We use a 64-bit Murmur3 hash~\cite{mumur} to randomize the 32-bit input data. Murmur3 is a non-cryptographic hash function that is simple to implement while guaranteeing a uniform distribution of hash values~\cite{frietag}. The math and logic (multiply and rotate) operations of the hash function are mapped to the dedicated Digital Signal Processing (DSP) slices~\cite{ug579} of the FPGA. A DSP slice contains pipeline registers to enhance the speed and efficiency of the application. The data flow engine takes advantage of the parallelism offered by the reconfigurable logic by inferring multiple DSP resources and schedules the math operations in a pipeline structure. 

\subsubsection{\textbf{Index Extractor}}
The index extractor module is fed with the 64-bit hash values. It extracts the first 16-bits of the hash as an index to identify the corresponding counter. The remaining 48-bits are forwarded to the leading zero detector.

\subsubsection{\textbf{Leading Zero Detector}}
This module determines and forwards the count of leading zeros. It leverages the efficiently synthesizable \texttt{CountLeadingZero} member function of the variable-width integer data type \texttt{ap\_uint<N>} provided by the Vivado HLS libraries.

\subsubsection{\textbf{Buckets}}
The bucket counters are mapped onto dual-port BRAM~\cite{ug573} modules of the FPGA. The counters maintain the maximum ranks encountered for their respective buckets. The potential counter update is pipelined itself, first (a) reading the counter value identified by the extracted index, then (b) comparing it to the new computed leading zero count, and, finally, (c) updating the stored maximum if a larger rank has been encountered. Updates to the same counter that arrive during this read-modify-write cycle are merged.

\subsubsection{\textbf{Zero Counter and Bypass}}
This is a pass-through module forwarding the aggregated ranks to the harmonic mean computation. While doing so, it determines the number $V$ of counters that have remained unchanged at a value of zero.

\subsubsection{\textbf{Harmonic Mean}}
This module computes the harmonic mean of the aggregated ranks. Observe that the corresponding summation kernel accumulates powers of two, particularly $2^{-M[j]}$. Thus, these addends can be formed easily from a 1-hot code asserting the corresponding binary fractional bit. They are accumulated onto an arbitrary-precision fixed-point data type of the HLS library \cite{vhls} with $m$ binary integer digits and $H+p+1$ binary fractional digits to attain an exact sum. After processing all aggregated ranks, the raw cardinality estimate $E$ is computed using HLS-synthesized floating-point arithmetic.

\subsubsection{\textbf{Correction}}
The correction module replaces small raw HLL cardinality estimates $E$ by estimates obtained through linear counting. This small range correction is triggered when the raw estimate falls below the threshold $E\le\frac{5}{2}m$ and empty buckets have been observed, i.e. $V\ne0$. The number of empty buckets serves as an input to the linear counting computation. Using a 64-bit hash function, a large range correction is not required. The final cardinality estimate achieves an accuracy with a typical standard error below 2\%.

\subsection{Parallel Architecture}\label{sec:parallel}
As shown in \rfig{fig:hll_parallel}, the HLL aggregation phase can be trivially parallelized into $k$ \emph{independent} but otherwise identical aggregation pipelines. This allows to scale the input bandwidth of the computation perfectly to $k\times$ $32$-bit words per cycle at the cost of additional FPGA resources. The input data is simply sliced to feed the individual pipelines. Slicing the multi-word input only implies wiring for the actual data and minimal handshaking with the pipeline inputs. Inputs are processed where they arrive with no active reassignment to particular pipelines. After aggregation, the partial sketches of each pipeline are merged by taking the maximum rank across corresponding buckets by the Merge buckets module. Its complexity is that of a fold. The partial sketches are streamed in parallel and folded bucket by bucket. The following computation phase is identical to the one used in the non-parallel architecture.
\section{Experiments and Results}\label{sec:exp}
The HLL design is coupled with a Xilinx XDMA bridge IP (a subsystem for PCIe 3.0 $\times 16$)~\cite{pg195} on a Xilinx Virtex UltraScale+ FPGA (VCU118) platform~\cite{ug1224} serving as a PCIe endpoint for external communication.

The HLL design is driven by 322 MHz (with time period 3.1 ns) clock provided by the CMAC module of the Xilinx UltraScale+ 100G Ethernet subsystem\mbox{~\cite{pg203}}. Thus, with $II = 1$ each pipeline operates at a throughput of \mbox{322\,MHz$\times$32\,bits = 10.3 Gbit/s.}

\subsection{Throughput}
Increasing the number of pipelines in the design allows to scale the throughput until saturating the PCIe bandwidth. This dependency is shown by \rfig{fig:fpga_vs_pip}. The figure contrasts the practically measured throughput with the theoretical throughput obtained by aggregating the processing rate across all pipelines. Both graphs show the same linear growth up to 10~parallel pipelines. At this point, the PCIe bandwidth is saturated \mbox{(10$\times$10.3\,Gbit/s = 103\,Gbit/s~>\,12.48 GByte/s)}. Adding more pipelines can no longer boost the throughput as the design is I/O bounded (PCIe bound). Thus, for the rest of the PCIe-dependent experiments, we scale the system to at most 10~parallel pipelines.

\subsection{HLL Standard Error}
We validated our implementation with same data sets as used for the profiling described in \rsec{sec:profiling}. The standard error of the FPGA-implemented HLL with $p=16$ and using a $64$-bit hash matches the standard error curve $HLL64$ in \rfig{fig:p=16}.

\subsection{CPU Performance}
\begin{figure}[t]
  \begin{subfigure}[b]{0.5\linewidth}
    \vspace*{-6pt}
    \includegraphics[width=\linewidth]{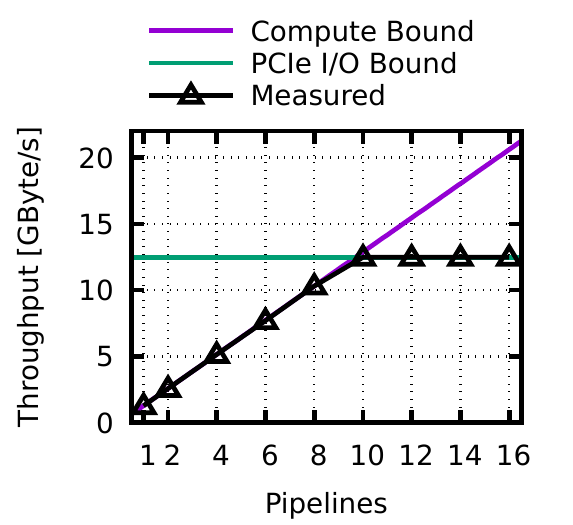}
    \vspace*{-12pt}
    \caption{\scriptsize FPGA throughput vs. \#pipelines.}
    \label{fig:fpga_vs_pip} 
  \end{subfigure}%
  \begin{subfigure}[b]{0.5\linewidth}
    \vspace*{-6pt}
    \includegraphics[width=\linewidth]{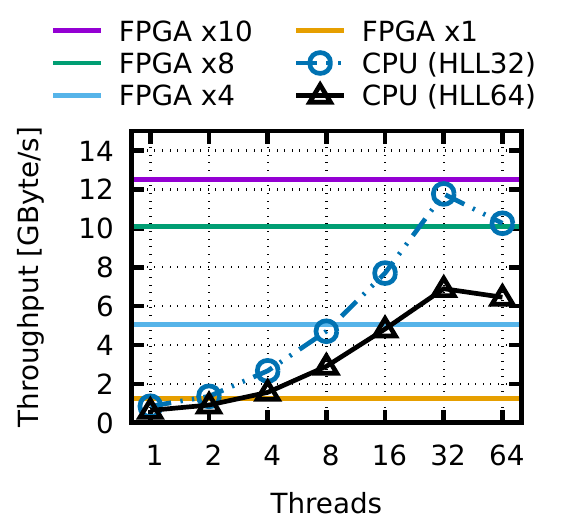}
    \vspace*{-12pt}
    \caption{\scriptsize CPU throughput vs. \#threads.}
    \label{fig:cpu_vs_thread} 
  \end{subfigure}
 \caption{HyperLogLog throughput.}
 \vspace*{-12pt}
 \label{fig:throughput} 
\end{figure}

As a baseline, we have implemented HLL using both a 32-bit hash and a 64-bit hash in C++. The 32-bit Murmur3 implementation was optimized using AVX2 technology \cite{avx2} leveraging, 8-fold vectorization parallelism. The corresponding 4-fold vectorization of the 64-bit hash did not prove beneficial as there is no native $64\times64$ bit vector multiplication instruction in the AVX2 instruction set. In both cases, the leading zero detection exploits GCC's \texttt{\_\_builtin\_clz}, which maps favorably to the native \texttt{x86} instruction \texttt{LZCNT} that directly implements the desired operation. Threads are used to parallelize the aggregation phase on a dual-socket Intel\,\textsuperscript{\tiny\textregistered} Xeon\,\textsuperscript{\tiny\textregistered} E5-2630~v3 system with a total of 16 cores, clocked at 2.40\,GHz.

\rfig{fig:fpga_vs_pip} shows that the performance of the FPGA implementation scales perfectly with the number of parallel pipelines until encountering its I/O bound, which is exhausted with 10 pipelines. Note that the achieved throughput is identical for the 32-bit and the 64-bit hash as the extra effort required for the more capable hash function is unrolled.

\rfig{fig:cpu_vs_thread} shows the analogous performance scaling by increasing the thread-level parallelism of the CPU implementations. It halts and even slightly reverses when the number of forked threads exceeds the native support on the system. The choice of the hash function has a direct impact on the observable performance. The use of the 64-bit hash, which enables the support for cardinalities beyond $10^8$, reduces the performance to about 60\% of the one achieved for the 32-bit hash function. In either case, the fully unrolled FPGA implementation outperforms even the fully designated dual-processor system, for the 64-bit hash by more than 80\%. As the FPGA-based Murmur3 implementation can be unrolled in space to a pipeline with $II=1$, it achieves a significant speedup over a CPU implementation~\cite{hash_accel}. The CPU implementation of the HLL is strictly compute-bound by the hash computation even after using AVX2 extensions. The FPGA implementation benefits from the dataflow architecture and the parallel implementation of the hash function. The FPGA-based HLL takes advantage of dataflow architecture that enables to estimate the statistics while data is streamed. The data and statistics arrive almost together in contrast to the CPU implementation.
\subsection{FPGA Resource Utilization}

\begin{table}
\caption{Resources usage of HLL vs. \#Pipelines.}
\resizebox{\linewidth}{!}{%
\begin{tabular}{l@{}r@{\,/\,}rr@{\,/\,}rr@{\,/\,}rr@{\,/\,}rr@{\,/\,}rr@{\,/\,}r}\toprule
\textbf{Pipelines} & \multicolumn{2}{c}{1} & \multicolumn{2}{c}{2} & \multicolumn{2}{c}{4} & \multicolumn{2}{c}{8} & \multicolumn{2}{c}{10} & \multicolumn{2}{c}{16} \\\midrule
\textbf{BRAM} & 12 & 0.55\% 
              & 24 & 1.11\% 
              & 48 & 2.22\% 
              & 96 & 4.44\% 
              & 120 & 5.55\%    
              & 192 & 8.88\%\\
\textbf{DSP}  & 84 & 1.22\% 
              & 152 & 2.22\%
              & 288 & 4.21\% 
              & 560 & 8.18\%  
              & 696 & 10.18\%   
              & 1104 & 16.14\%\\
\textbf{LUT}  & 4.5K & 0.38\%    
              & 5.5K & 0.46\%    
              & 7.3K & 0.62\%    
              & 11.2K & 0.95\%   
              & 13.1K & 1.10\%   
              & 18.9K & 1.60\%\\ 
\textbf{FF}   & 5.5K & 0.23\%
              & 6.9K & 0.29\% 
              & 9.5K & 0.40\% 
              & 15.4K & 0.65\%  
              & 18.3K & 0.77\%  
              & 26.8K & 1.13\%\\
\bottomrule
\multicolumn{13}{l}{\footnotesize HLL resource utilization for HLL64 and p=16 on a XCVU9P device (VCU118 Board)}
\end{tabular}}
\label{tab:resources}
\end{table}

While HLL itself possess a modest memory requirement that grows logarithmically, with the used hash size and $k$ independent pipelines result in FPGA resources utilization shown in \rtab{tab:resources}. The table summarizes the resource utilization for the FPGA-based HLL for 32- and 64-bit hashes, each with precision $p=16$. Clearly, the resource utilization scales linearly with the number of pipelines. It is the investment of these resources that pay for the gained throughput. 

The LUTs and FFs utilization remain under 2\%, thus exhibiting its lightweight property in terms of logic resource consumption. Since each pipeline has a dedicated counter-memory, an increase in the number of pipelines is reflected in the BRAM utilization. The maximum usage of this critical resource remains under 6\%. The other critical resource is the DSP blocks that are mostly consumed by the hash computation. Slightly more than 10\% are consumed by 10~pipelines using a 64-bit hash implementation. On the given device, this resource type would eventually limit further scaling.
\section{HyperLogLog on TCP/IP Networks}\label{sec:tcp}
\begin{figure}[b]
\centering
\includegraphics[width=0.71\linewidth]{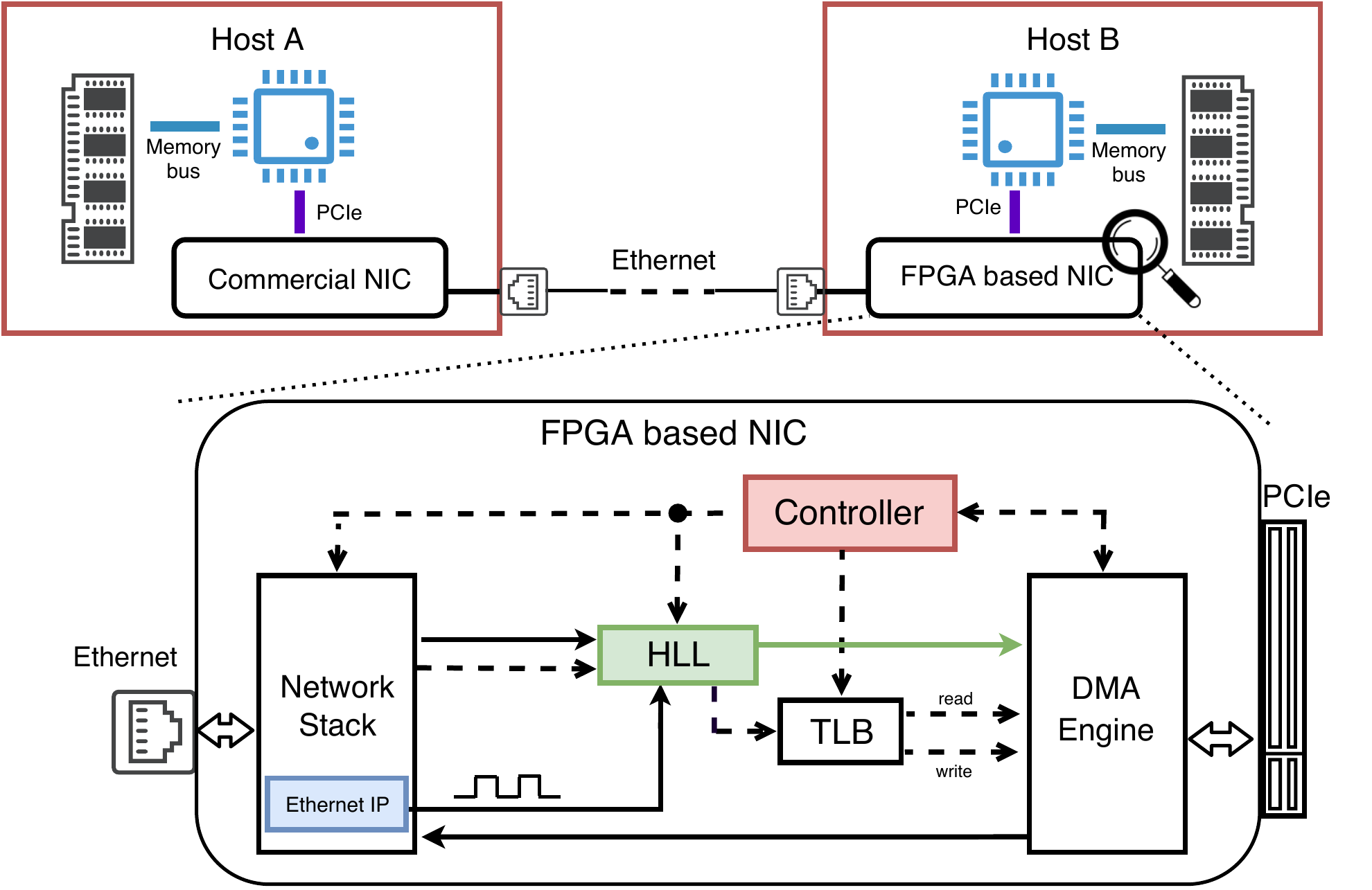}
\caption{HyperLogLog on an FPGA-based NIC.}
\vspace{-15pt}
\label{fig:hll_sys}
\end{figure}

The HLL design is deployed in an FPGA-based NIC featuring a 100\,Gbit/s TCP/IP stack \cite{Ruiz19}. We show that the implementation can process incoming data at line rate.

\rfig{fig:hll_sys} depicts the system-level design used to deploy the HLL in the network. The two hosts, A and B, are attached to the network via dedicated PCIe-attached NICs. Host~A uses the commercial NIC Mellanox ConnectX-5~\cite{mellanox} across PCIe Gen\,3.0 $\times16$. Host~B uses an FPGA-based NIC, which includes a network stack using the UltraScale+ 100G Ethernet Subsystem for a 100\,Gbit/s network. This network stack and the integrated HLL implementation are managed by a controller module.

The design is divided into two clock domains: (a) the PCIe clock domain at 250\,MHz and (b) the network clock domain established by the serial transceiver interface at 322\,MHz for the 100\,Gbit/s network. Since the HLL module is processing the data from the network, it is placed in the network clock domain.

We validate the HLL integration on the FPGA-based NIC and measure its sustained throughput for different numbers of parallel pipelines. Data is received over the network from Host~A. The network stack on the FPGA-based NIC extracts the data from the received packets to feed the HLL estimation. After processing the whole data stream, the estimation result is sent to Host~B's.
Once all the data is streamed, the time taken to compute the cardinality result remains constant, 203$\mu s$, irrespective of the quantity of data, and it stems mainly from the time taken to read all the contents from the counter buckets $(2^{16}\times3.1ns)$.

The throughput sustained by the receiving NIC is tabulated in \rtab{tab:nic_throughput}. For one and two parallel pipelines, the integrated HLL processing induces significant back-pressure on the network stack, which starts dropping packets. The resulting packet re-transmission cycles push the observable throughput way below 1\,Gbit/s. Scaling to more parallel pipelines allows the flow control to work effectively and enables a steady growth of the sustainable throughput. Indeed, four pipelines would be sufficient to saturate 10 and 25\,Gbit/s networks, and eight pipelines wold be able to handle a 40\,Gbit/s network stream safely. In order to accommodate a 100\,Gbit/s network, without inducing any back pressure on it, 16 pipelines are needed. The increase in the number of pipelines from 10, sustaining a 12\,GByte/s PCIe throughput, to 16, sustaining a 9.35\,GByte/s network throughput, comes as a result of supporting network's bursty behaviour.  

As the HLL throughput can be scaled by adding more pipelines and by increasing clock rates, this solution will continue to benefit from the technological improvements of future devices. Currently, HLL on a NIC can achieve a 35\% higher processing rate than a 16-core CPU for the same statistical guarantees (\rfig{fig:cpu_vs_thread}).
This gives an idea of the potential of our HLL design in improving computational efficiency in cloud settings.
\section{Conclusions}\label{sec:conc}

\begin{table}[t]
\centering
\caption{Throughput [GByte/s] vs. \#Pipelines.}
\label{tab:nic_throughput}
\begin{tabular}{lccccccc}\toprule
Pipelines & 1 & 2 & 4 & 8 & 10 & 16 \\\midrule
Throughput & 0.05 & 0.12 & 4.83 & 6.77 & 8.94 & 9.35 \\\bottomrule
\end{tabular}
\end{table}
We presented how the FPGA-based custom hardware implementation of HyperLogLog can outperform a modern multi-core CPU system. We exploit the programmable logic of the FPGA to implement a parallelized, multi-pipelined HLL. This implementation scales throughput linearly with the number of pipelines, providing an adaptive solution that can be used in a variety of settings (near-storage if the stream comes from NVM or SSD, on the network data path, or near memory). In the paper, we illustrated this flexibility by discussing two deployments in detail: (a) an FPGA configured as a co-processor and (b) an FPGA functioning as a NIC. Since the standard error of HLL depends mainly on the quality of the hash function but the size of the hash function affects performance, we have provided a detailed analysis of error rates and implementation trade-offs on both the FPGA and the CPU. While on the FPGA, the bigger hash function results in a more intense utilization of resources; on the CPU, it leads to computation overhead that severely limits the processing rate. Nevertheless, even with more extensive hash functions, we can still place enough pipelines on the FPGA to saturate the system on its I/O bounds. We see the results as a promising first step to offload the essential operations such as HyperLogLog to FPGA-based accelerators as the performance difference over CPUs can be leveraged to reduce the overall number of machines needed to process extensive data collections without compromising either performance nor result quality.
\bibliographystyle{IEEEtran}
\bibliography{mybib}

\begin{thebibliography}{10}
\providecommand{\url}[1]{#1}
\csname url@samestyle\endcsname
\providecommand{\newblock}{\relax}
\providecommand{\bibinfo}[2]{#2}
\providecommand{\BIBentrySTDinterwordspacing}{\spaceskip=0pt\relax}
\providecommand{\BIBentryALTinterwordstretchfactor}{4}
\providecommand{\BIBentryALTinterwordspacing}{\spaceskip=\fontdimen2\font plus
\BIBentryALTinterwordstretchfactor\fontdimen3\font minus
  \fontdimen4\font\relax}
\providecommand{\BIBforeignlanguage}[2]{{%
\expandafter\ifx\csname l@#1\endcsname\relax
\typeout{** WARNING: IEEEtran.bst: No hyphenation pattern has been}%
\typeout{** loaded for the language `#1'. Using the pattern for}%
\typeout{** the default language instead.}%
\else
\language=\csname l@#1\endcsname
\fi
#2}}
\providecommand{\BIBdecl}{\relax}
\BIBdecl

\bibitem{freq_item}
J.~{Teubner}, R.~{Muller}, and G.~{Alonso}, ``Frequent item computation on a
  chip,'' \emph{IEEE Transactions on Knowledge and Data Engineering}, vol.~23,
  no.~8, pp. 1169--1181, Aug 2011.

\bibitem{cormode}
\BIBentryALTinterwordspacing
G.~Cormode, ``Data sketching,'' \emph{Communications of the ACM}, vol.~60,
  no.~9, pp. 48--55, 2017. [Online]. Available:
  \url{https://doi.org/10.1145/3080008}
\BIBentrySTDinterwordspacing

\bibitem{heule}
\BIBentryALTinterwordspacing
S.~Heule, M.~Nunkesser, and A.~Hall, ``{HyperLogLog in practice: algorithmic
  engineering of a state of the art cardinality estimation algorithm},'' in
  \emph{EDBT}, 2013, pp. 683--692. [Online]. Available:
  \url{http://doi.acm.org/10.1145/2452376.2452456}
\BIBentrySTDinterwordspacing

\bibitem{flajolet}
P.~Flajolet, \'{E}ric Fusy, O.~Gandouet, and F.~Meunier, ``{HyperLogLog: The
  analysis of a near-optimal cardinality estimation algorithm},'' in
  \emph{{AOFA}}, 2007.

\bibitem{google_blog}
\BIBentryALTinterwordspacing
F.~Hoffa, ``{Counting uniques faster in BigQuery with HyperLogLog++},'' 2017,
  last accessed September 2019. [Online]. Available:
  \url{https://cloud.google.com/blog/products/gcp/counting-uniques-faster-in-bigquery-with-hyperloglog}
\BIBentrySTDinterwordspacing

\bibitem{chaudhuri2017approximate}
S.~Chaudhuri, B.~Ding, and S.~Kandula, ``Approximate query processing: No
  silver bullet,'' in \emph{SIGMOD}, 2017, pp. 511--519.

\bibitem{youssefi1979query}
K.~Youssefi and E.~Wong, ``Query processing in a relational database management
  system,'' in \emph{VLDB}, 1979, pp. 409--417.

\bibitem{zhang2009kernel}
Z.~Zhang, Y.~Yang, R.~Cai, D.~Papadias, and A.~Tung, ``Kernel-based skyline
  cardinality estimation,'' in \emph{SIGMOD}, 2009, pp. 509--522.

\bibitem{neumann2011characteristic}
T.~Neumann and G.~Moerkotte, ``Characteristic sets: Accurate cardinality
  estimation for rdf queries with multiple joins,'' in \emph{ICDE}, 2011, pp.
  984--994.

\bibitem{estan2003bitmap}
C.~Estan, G.~Varghese, and M.~Fisk, ``Bitmap algorithms for counting active
  flows on high speed links,'' in \emph{SIGCOMM}, 2003, pp. 153--166.

\bibitem{varagnolo2013distributed}
D.~Varagnolo, G.~Pillonetto, and L.~Schenato, ``Distributed cardinality
  estimation in anonymous networks,'' \emph{IEEE Transactions on Automatic
  Control}, vol.~59, no.~3, pp. 645--659, 2013.

\bibitem{metwally2008go}
A.~Metwally, D.~Agrawal, and A.~E. Abbadi, ``Why go logarithmic if we can go
  linear?: Towards effective distinct counting of search traffic,'' in
  \emph{EDBT}, 2008, pp. 618--629.

\bibitem{eeckhout2017moore}
L.~Eeckhout, ``{Is Moore's Law Slowing Down? What's Next?}'' \emph{IEEE Micro},
  no.~4, pp. 4--5, 2017.

\bibitem{putnam2014reconfigurable}
A.~Putnam, A.~M. Caulfield, E.~S. Chung, D.~Chiou, K.~Constantinides, J.~Demme,
  H.~Esmaeilzadeh, J.~Fowers, G.~P. Gopal, J.~Gray \emph{et~al.}, ``A
  reconfigurable fabric for accelerating large-scale datacenter services,'' in
  \emph{ISCA}, 2014, pp. 13--24.

\bibitem{amazonf1}
\BIBentryALTinterwordspacing
J.~Barr, ``{Amazon F1 Instances},'' 2019, last accessed January 2020. [Online].
  Available: \url{https://aws.amazon.com/ec2/instance-types/f1}
\BIBentrySTDinterwordspacing

\bibitem{jouppi2017datacenter}
N.~P. Jouppi, C.~Young, N.~Patil, D.~Patterson, G.~Agrawal, R.~Bajwa, S.~Bates,
  S.~Bhatia, N.~Boden, A.~Borchers \emph{et~al.}, ``In-datacenter performance
  analysis of a tensor processing unit,'' in \emph{ISCA}, 2017, pp. 1--12.

\bibitem{oliver2011reconfigurable}
N.~Oliver, R.~R. Sharma, S.~Chang, B.~Chitlur, E.~Garcia, J.~Grecco, A.~Grier,
  N.~Ijih, Y.~Liu, P.~Marolia \emph{et~al.}, ``{A reconfigurable computing
  system based on a cache-coherent fabric},'' in \emph{ReConFig}, 2011, pp.
  80--85.

\bibitem{halstead2013accelerating}
R.~J. Halstead, B.~Sukhwani, H.~Min, M.~Thoennes, P.~Dube, S.~Asaad, and
  B.~Iyer, ``{Accelerating join operation for relational databases with
  FPGAs},'' in \emph{FCCM}, 2013, pp. 17--20.

\bibitem{kara2017fpga}
K.~Kara, J.~Giceva, and G.~Alonso, ``{FPGA-based Data Partitioning},'' in
  \emph{SIGMOD}, 2017, pp. 433--445.

\bibitem{mueller2009data}
R.~Mueller, J.~Teubner, and G.~Alonso, ``{Data processing on FPGAs},''
  \emph{VLDB Endowment}, vol.~2, no.~1, pp. 910--921, 2009.

\bibitem{mashimo2017high}
S.~Mashimo, T.~Van~Chu, and K.~Kise, ``High-performance hardware merge
  sorter,'' in \emph{FCCM}, 2017, pp. 1--8.

\bibitem{istvan2014histograms}
Z.~Istvan, L.~Woods, and G.~Alonso, ``Histograms as a side effect of data
  movement for big data,'' in \emph{SIGMOD}, 2014, pp. 1567--1578.

\bibitem{hash_accel}
K.~Kara and G.~Alonso, ``Fast and robust hashing for database operators,'' in
  \emph{FPL}, 2016, pp. 1--4.

\bibitem{mc1}
\BIBentryALTinterwordspacing
D.~Tong, L.~Sun, K.~Matam, and V.~Prasanna, ``{High throughput and programmable
  online traffic classifier on FPGA},'' in \emph{FPGA}, 2013, pp. 255--264.
  [Online]. Available: \url{http://doi.acm.org/10.1145/2435264.2435307}
\BIBentrySTDinterwordspacing

\bibitem{mc11}
\BIBentryALTinterwordspacing
D.~Tong and V.~Prasanna, ``High throughput sketch based online heavy hitter
  detection on fpga,'' \emph{SIGARCH Comput. Archit. News}, vol.~43, no.~4, p.
  70–75, Apr. 2016. [Online]. Available:
  \url{https://doi.org/10.1145/2927964.2927977}
\BIBentrySTDinterwordspacing

\bibitem{sidler2015scalable}
D.~Sidler, G.~Alonso, M.~Blott, K.~Karras, K.~Vissers, and R.~Carley,
  ``{Scalable 10Gbps TCP/IP Stack Architecture for Reconfigurable Hardware},''
  in \emph{FCCM}.\hskip 1em plus 0.5em minus 0.4em\relax IEEE, 2015, pp.
  36--43.

\bibitem{sidlerrepo}
\BIBentryALTinterwordspacing
D.~Sidler, ``{Scalable Network Stack for FPGAs (TCP/IP, RoCEv2)},'' 2019, last
  accessed January 2020. [Online]. Available:
  \url{https://github.com/fpgasystems/fpga-network-stack}
\BIBentrySTDinterwordspacing

\bibitem{chung2018serving}
E.~Chung, J.~Fowers, K.~Ovtcharov, M.~Papamichael, A.~Caulfield, T.~Massengill,
  M.~Liu, D.~Lo, S.~Alkalay, M.~Haselman \emph{et~al.}, ``{Serving DNNs in Real
  Time at Datacenter Scale with Project Brainwave},'' \emph{IEEE Micro},
  vol.~38, no.~2, pp. 8--20, 2018.

\bibitem{stochastic}
P.~Clifford and I.~A. Cosma, ``A statistical analysis of probabilistic counting
  algorithms,'' \emph{Scandinavian Journal of Statistics}, vol.~39, no.~1, pp.
  1--14, 2012.

\bibitem{indyk}
\BIBentryALTinterwordspacing
P.~Indyk and D.~Woodruff, ``Tight lower bounds for the distinct elements
  problem,'' in \emph{FOCS}, 2003, pp. 283--288. [Online]. Available:
  \url{http://dl.acm.org/citation.cfm?id=946243.946312}
\BIBentrySTDinterwordspacing

\bibitem{Kane}
\BIBentryALTinterwordspacing
D.~M. Kane, J.~Nelson, and D.~P. Woodruff, ``An optimal algorithm for the
  distinct elements problem,'' in \emph{PODS}, 2010, pp. 41--52. [Online].
  Available: \url{http://doi.acm.org/10.1145/1807085.1807094}
\BIBentrySTDinterwordspacing

\bibitem{mumur}
\BIBentryALTinterwordspacing
Aappleby, ``{aappleby/SMHasher},'' 2016, last accessed January 2020. [Online].
  Available: \url{https://github.com/aappleby/smhasher}
\BIBentrySTDinterwordspacing

\bibitem{hls_book}
R.~Kastner, J.~Matai, and S.~Neuendorffer, ``{Parallel programming for
  FPGAs},'' \emph{arXiv preprint arXiv:1805.03648}, 2018.

\bibitem{frietag}
M.~J. Freitag and T.~Neumann, ``Every row counts: Combining sketches and
  sampling for accurate group-by result estimates,'' in \emph{CIDR}, 2019.

\bibitem{ug579}
\BIBentryALTinterwordspacing
{Xilinx}, ``{UltraScale Architecture DSP Slice User Guide (UG579, v1.9)},''
  2019, last accessed December 2019. [Online]. Available:
  \url{https://www.xilinx.com/support/documentation/user\_guides/ug579-ultrascale-dsp.pdf}
\BIBentrySTDinterwordspacing

\bibitem{ug573}
\BIBentryALTinterwordspacing
Xilinx, ``{UltraScale Architecture Memory Resources (UG573, v1.10)},'' 2019,
  last accessed September 2019. [Online]. Available:
  \url{https://www.xilinx.com/support/documentation/user\_guides/ug573-ultrascale-memory-resources.pdf}
\BIBentrySTDinterwordspacing

\bibitem{vhls}
\BIBentryALTinterwordspacing
{Xilinx}, ``{Vivado Design Suite User Guide High-Level Synthesis UG902
  (v2019.1)},'' 2019, last accessed September 2019. [Online]. Available:
  \url{https://www.xilinx.com/support/documentation/sw\_manuals/xilinx2019\_1/ug902-vivado-high-level-synthesis.pdf}
\BIBentrySTDinterwordspacing

\bibitem{pg195}
\BIBentryALTinterwordspacing
Xilinx, ``{DMA/Bridge Subsystem for PCI Express v4.1 PG195 (2019.1)},'' 2019,
  last accessed December 2019. [Online]. Available:
  \url{https://www.xilinx.com/support/documentation/ip\_documentation/xdma/v4\_1/pg195-pcie-dma.pdf}
\BIBentrySTDinterwordspacing

\bibitem{ug1224}
\BIBentryALTinterwordspacing
{Xilinx}, ``{VCU118 Evaluation Board (UG1224, v1.4)},'' 2018, last accessed
  December 2019. [Online]. Available:
  \url{https://www.xilinx.com/support/documentation/boards\_and\_kits/vcu118/ug1224-vcu118-eval-bd.pdf}
\BIBentrySTDinterwordspacing

\bibitem{pg203}
\BIBentryALTinterwordspacing
Xilinx, ``{UltraScale+ Devices Integrated 100G Ethernet Subsystem (v2.4)},''
  2018, last accessed September 2019. [Online]. Available:
  \url{https://www.xilinx.com/support/documentation/ip\_documentation/cmac\_usplus/v2\_4/pg203-cmac-usplus.pdf}
\BIBentrySTDinterwordspacing

\bibitem{avx2}
P.~Gepner, ``{Using AVX2 instruction set to increase performance of high
  performance computing code},'' \emph{Computing and Informatics}, vol.~36,
  no.~5, pp. 1001--1018, 2017.

\bibitem{Ruiz19}
M.~Ruiz, D.~Sidler, G.~Sutter, G.~Alonso, and S.~L{\'{o}}pez{-}Buedo,
  ``{Limago: An FPGA-Based Open-Source 100 GbE {TCP/IP} Stack},'' in
  \emph{FPL}, 2019, pp. 286--292.

\bibitem{mellanox}
\BIBentryALTinterwordspacing
Mellanox, ``{ConnectX-5 EN Card, Product Brief},'' 2019, last accessed
  September 2019. [Online]. Available:
  \url{https://www.mellanox.com/related-docs/prod\_adapter\_cards/PB\_ConnectX-5\_EN\_Card.pdf}
\BIBentrySTDinterwordspacing

\end{thebibliography}
\end{document}